\documentclass[aps,twocolumn,groupedaddress]{revtex4}
\usepackage{amsmath}
\usepackage{graphics,epsfig,graphicx,epstopdf}

\DeclareGraphicsExtensions{.pdf}

\newcommand {\vek}[1]{\mathbf{#1}}

\begin{document}

\title{The role of input noise in transcriptional regulation}

\author{Ga\v{s}per Tka\v{c}ik$^a$, Thomas Gregor$^{a,b}$ and William Bialek$^{a,c}$}

\affiliation{$^a$Joseph Henry Laboratories of Physics, Lewis--Sigler Institute for Integrative Genomics,  $^b$Howard Hughes Medical Institute
and $^c$Princeton Center for Theoretical Physics,
Princeton University, Princeton, New Jersey 08544}

\begin{abstract}
Even under constant external conditions, the expression levels of genes fluctuate. Much emphasis has been placed on the components of this noise that are due to randomness in  transcription and translation; here we analyze the role of noise associated with the inputs to transcriptional regulation, the random arrival and binding of transcription factors to their target sites along the genome.  This noise sets a fundamental physical limit to the reliability of genetic control, and has clear signatures, but we show that these are easily obscured by experimental limitations and even by conventional methods for plotting the variance vs. mean expression level.  We argue that simple, global models of noise dominated by transcription and translation are inconsistent with the embedding of gene expression in a network of regulatory interactions.
Analysis of recent experiments on transcriptional control in the early {\em Drosophila} embryo shows that these results are quantitatively consistent with the predicted signatures of input noise, and we discuss the experiments needed to test the importance of input noise more generally.
\end{abstract}

\date{\today}
\maketitle

\section{Introduction}

A number of recent experiments have focused attention on noise in  gene expression
\cite{elowitz+al_02,ozbudak+al_02,blake+al_03,raser+oshea_04,rosenfeld+al_05,pedraza+oudenaarden_05,golding+al_05,newman+al_06,bar-even+al_06}.  The study of noise in biological systems more generally has a long history, with two very different streams of thought.  On the one hand, observations of noise in behavior at the cellular or even organismal level give us a window into mechanisms at a much more microscopic level.  The classic example of using noise to draw inferences about biological mechanism is perhaps the Luria--Delbr\"uck experiment \cite{luria+delbruck_43}, which demonstrated the random character of mutations, but one can also point to early work on the nature of chemical transmission at synapses \cite{fatt+katz_50, fatt+katz_52} and on the dynamics of ion channel proteins \cite{lecar+nossal_71a,lecar+nossal_71b,stevens_72,conti+al_75}.  On the other hand, noise limits the reliability of biological function, and it is important to identify these limits.  Examples include tracking the reliability of visual perception at low light levels down to the ability of the visual system to count single photons \cite{hecht+al_42,barlow_81}, the implications of channel noise for the reliability of neural coding \cite{verveen+derksen_65,verveen+derksen_68,schneidman+al_98}, and the approach of bacterial chemotactic performance to the limits set by the random arrival of individual molecules at the cell surface \cite{berg+purcell_77}.

After demonstrating that one can observe noise in gene expression, most investigators have concentrated on the mechanistic implications of this noise.  Working backward from the observation of protein concentrations, one can try to find the components of noise that derive from the translation of messenger RNA into protein, or the components that arise from noise in the transcription and degradation of the mRNA itself.  At least in some organisms, a single mRNA transcript can give rise to many protein molecules, and this `burst' both amplifies the fluctuations in mRNA copy number and changes their statistics, so that even if the number of mRNA copies obeys the Poisson distribution the number of protein molecules will not \cite{paulsson_04}.   This discussion parallels the understanding that Poisson arrival of photons at the retina generates non--Poisson statistics of action potentials in retinal ganglion cells because each photon triggers a burst of spikes \cite{barlow+al_71}.  Recent large scale surveys of noise in eukaryotic transcription have suggested that the noise in most protein levels can be understood in terms of this picture, so that the fractional variance in the number of proteins $p_{\rm i}$ expressed from gene $\rm i$ is given by
\begin{equation}
\eta_{\rm i}^2 \equiv {{\langle (\delta p_{\rm i} )^2\rangle}\over{\langle p_{\rm i} \rangle^2}} = {b\over{\langle p_{\rm i} \rangle}},
\label{global_model}
\end{equation}
where $b\sim 10^3$ is the burst size, and is approximately constant for all genes
\cite{bar-even+al_06}.  

\begin{figure}[b]
   \centering
\epsfig{file=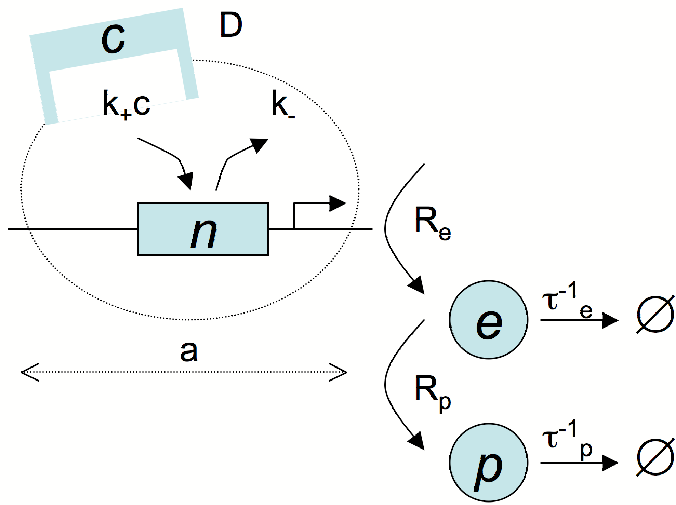,width=3.5in}
   \caption{A simple model for transcriptional regulation. Transcription factor is present at an average concentration $c$, diffusing freely with diffusion constant $D$; it can bind to the binding site of linear dimension $a$ and the fractional occupancy of this site is $n\in [0,1]$. Binding occurs with a second order rate constant $k_+$, and unbinding occurs with a first order rate constant $k_-$.  When the site is bound, the mRNA  are transcribed at rate $R_e$ and degraded with rate $\tau_e^{-1}$, resulting in a number of transcripts $e$. Proteins are translated from each mRNA molecule with rate $R_p$ and degraded with rate $\tau_p^{-1}$, resulting in a copy number $p$.}   
\label{f-model}
\end{figure}

The mechanistic focus on noise in transcription vs translation perhaps misses the functional role of gene expression as part of a regulatory network.  Almost all genes are subject to transcriptional regulation, and hence the expression level of a particular protein can be viewed as the cell's response to the concentration of the relevant transcription factors.   
Seen in this way, transcription and translation are at the `output' side of the response, and the binding of transcription factors to their targets along the genome is at the `input' side (Fig \ref{f-model}).  Noise can arise at both the input and output, and while fluctuations in transcription factor concentration could be viewed as an extrinsic source of noise \cite{elowitz+al_02,swain+al_02}, there will be fluctuations in target site occupancy even at fixed transcription factor concentration \cite{bialek+setayeshgar_05,walczak+al_05,zon+al_06}.  There is a physical limit to how much the impact of these input fluctuations can be reduced, essentially because any physical device that responds to changes in concentration is limited by shot noise in the diffusive arrival of the relevant molecules at their target sites \cite{bialek+setayeshgar_05,berg+purcell_77,bialek+setayeshgar_06}.

In this paper we revisit the relative contributions of input and output noise.  Input noise has a clear signature, namely that its impact on the output protein concentration peaks at an intermediate value of the input transcription factor concentration.  The analogous signature was essential, for example, in identifying the noise from random opening and closing of individual ion channels in neurons \cite{sigworth_77,sigworth_80}.  Perhaps surprisingly, we show that this signature is easily obscured in conventional ways of plotting the data on noise in gene expression.    Recent experiments on the regulation of Hunchback expression  by Bicoid in the early {\em Drosophila} embryo \cite{gregor_05,gregor+al_06b} are consistent with the predicted signature of input noise, and (although there are caveats) a quantitative analysis of these data supports a dominant contribution of diffusive shot noise.  We discuss what experiments would be required to test this conclusion more generally.  We begin, however, by asking whether any simple global model such as 
Eq (\ref{global_model}) can be consistent with the imbedding of gene expression in a network of regulatory interactions.

\section{Global consistency?}

Consider a gene $\rm i$ which is regulated by several transcription factors.  In steady state, the mean number of these proteins in the cell will be a function of the copy numbers of all the relevant transcription factors:
\begin{equation}
\langle p_{\rm i} \rangle = g_{\rm i} (p_1 ,p_2 , \cdots , p_{\rm K})
\end{equation}
If the copy numbers of the transcription factors fluctuate, this noise will propagate through the input/output relation $g$ \cite{pedraza+oudenaarden_05,hooshangi+al_05}, so that
\begin{equation}
\langle (\delta p_{\rm i})^2 \rangle = \sum_{\mu =1}^{\rm K} \sum_{\nu =1}^{\rm K} 
{{\partial g_{\rm i}}\over{\partial p_\mu}}
{{\partial g_{\rm i}}\over { \partial p_\nu}} \langle \delta p_\mu \delta p_\nu \rangle 
+ \langle (\delta p_{\rm i})^2 \rangle_{\rm int} ,
\label{prop1}
\end{equation}
where we include the intrinsic noise $\langle (\delta p_{\rm i})^2 \rangle_{\rm int}$ that occurs at fixed transcription factor levels.

If the noise in gene expression is dominated by the processes of transcription and translation, and if the transcription factors are not regulating each other, then the correlations between fluctuations in the copy numbers of different proteins will be very small, so we expect that
\begin{equation}
\langle \delta p_\mu \delta p_\nu \rangle = \delta_{\mu\nu} \langle (\delta p_\mu )^2 \rangle .
\end{equation}
This allows us to simplify the propagation of noise in Eq (\ref{prop1}) to give
\begin{equation}
\langle (\delta p_{\rm i})^2 \rangle = \sum_{\mu =1}^{\rm K} 
\left({{\partial g_{\rm i}}\over{\partial p_\mu}}\right)^2
\langle (\delta p_\mu )^2 \rangle 
+ \langle (\delta p_{\rm i})^2 \rangle_{\rm int} .
\label{prop2}
\end{equation}
If, as in Eq (\ref{global_model}), we express the noise in protein copy number as a fractional noise $\eta$, then this becomes
\begin{equation}
\eta_{\rm i}^2 = \sum_{\mu =1}^{\rm K} 
\left({{\partial \log g_{\rm i}}\over{\partial \log p_\mu}}\right)^2
\eta_\mu^2
+\eta_{\rm i , int}^2  .
\label{prop3}
\end{equation}
In particular, this means that there is a minimum level of noise,
\begin{equation}
\eta_{\rm i}^2 \geq \sum_{\mu =1}^{\rm K} 
\left({{\partial \log g_{\rm i}}\over{\partial \log p_\mu}}\right)^2
\eta_\mu^2 .
\label{prop4}
\end{equation}
But if the fractional variance in protein copy number has a simple, global relation to the mean copy number, as in Eq (\ref{global_model}) \cite{bar-even+al_06}, then this simplifies still further:
\begin{eqnarray}
{b\over{\langle p_{\rm i}\rangle}} &\geq& \sum_{\mu =1}^{\rm K} 
\left({{\partial \log g_{\rm i}}\over{\partial \log p_\mu}}\right)^2
{b\over{\langle p_\mu \rangle}} \\
\Rightarrow 1 &\geq& \sum_{\mu =1}^{\rm K} 
\left({{\partial \log g_{\rm i}}\over{\partial \log p_\mu}}\right)^2
{{\langle p_{\rm i}\rangle } \over{\langle p_\mu \rangle}} .
\label{inequality}
\end{eqnarray}

Since the proteins labeled by the indices $\mu$ represent transcription factors, usually present at low concentrations, and the protein $\rm i$ is a regulated gene---such as a structural or metabolic protein---but not a transcription factor itself, one expects that $\langle p_{\rm i}\rangle /\langle p_\mu\rangle \gg 1$.  But then we have
\begin{equation}
\sum_{\mu =1}^{\rm K} 
\left({{\partial \log g_{\rm i}}\over{\partial \log p_\mu}}\right)^2
\ll 1 .
\label{inequality2}
\end{equation}
Since this inequality constrains the sum of squares of terms, each must be much smaller than one.  This means that when we make a small change the concentration of any transcription factor, the response of the regulated gene must be much less than proportional.  In this sense, the assumption of a simple global description for the level of noise in gene expression, Eq (\ref{global_model}), leads us to the conclusion that transcriptional ``regulation'' can't really be very effective, and this must be wrong.
Notice that this problem is independent of the burst size $b$, and hence doesn't depend on whether the noise is dominated by transcription or translation.   

Our conclusion from the inequality in Eq (\ref{inequality2}) is that we should re--examine the original hypothesis about noise, Eq (\ref{global_model}).  An alternative is that this hypothesis is correct, but that there are subtle correlations among all the protein copy number fluctuations of all the different transcription factors.  If we want the global output model to be correct, these correlations would have to take on a very special form---different transcription factors regulating a single gene would have to be correlated in  a way that matches their impact on the expression of that gene---which seems implausible but would be very interesting if it were true.  
 
\section{Sources of noise}

Figure \ref{f-model} makes clear that the concentration of a protein can fluctuate for many reasons.  The processes of synthesis and degradation of the protein molecules themselves are discrete and stochastic, as are the synthesis and degradation of mRNA molecules; together these constitute the ``output noise'' which has been widely discussed.  But if we are considering a gene whose transcription is regulated, we need a microscopic model for this process.  For the case of a transcriptional activator,  there are binding sites for the transcription factors upstream of the regulated gene, and when these sites are occupied transcription proceeds at some rate, but when the site is empty transcription is inhibited.  Because there are only a small number of relevant binding sites (in the simplest case, just one), the occupancy of these sites must fluctuate, and this random switching is an additional source of noise.  In addition, the binding of transcription factors to their target sites along the genome depends on the concentration in the immediate neighborhood of these sites, and this fluctuates as molecules diffuse into and out of the neighborhood.

All of the different processes described above and schematized in Fig \ref{f-model} can be analyzed analytically using Langevin methods, and the predictions of this analysis can be tested against detailed stochastic simulations.  The details of the analysis are given in Appendix A.  Notice that variations in cell size, protein sorting in cell division, fluctuations in RNA polymerase and ribosome concentrations, and all other extrinsic contributions to the noise are neglected.

When the dust settles, the variance in protein copy number $\sigma_{\rm p}^2$ can be written as a sum of three terms, which correspond to the output, switching, and diffusion noise.   To set the scale, we express the copy number as a fraction of its maximum possible mean value, $p_0$, which is reached at high concentrations of the transcriptional activator.  In these units, we find
\begin{equation}
\left(\frac{\sigma_{\rm p}}{p_0}\right)^2
=
\frac{1+R_p\tau_e}{p_0}\bar{p}
+
\frac{\left(1-\bar{p}\right)^2\bar{p}}{k_-\tau_p}
+
\frac{\left(1-\bar{p}\right)^2\bar{p}^2}{\pi D a {c}\tau_p} 
\label{eq-pnoise1}
\end{equation}
where ${\bar p} = \langle p \rangle/p_0$ is the protein copy number expressed as a fraction of its maximal value,  $c$ is the concentration of the transcription factor, and  other parameters are as explained in Fig \ref{f-model}.

The first term in Eq (\ref{eq-pnoise1}) is the output noise and has a Poisson--like behavior, with variance proportional to the mean, but the proportionality constant differs from 1 by $R_p\tau_e$, i.e. the burst size or the number of proteins produced per mRNA \cite{paulsson_04}. This is just the simple model of Eq (\ref{global_model}), with $b=1+R_p\tau_e$. 

The second term in Eq (\ref{eq-pnoise1}) originates from binomial ``switching''  as the transcription factor binding site occupation fluctuates, and is most closely analogous to the noise from random opening and closing of ion channels.  This term will be small for unbinding rates $k_-$ that are fast compared to the protein lifetime, but might be large for factors that take a long time to equilibrate or that form energetically stable complexes on their promoters.

The third term in Eq (\ref{eq-pnoise1}) arises because the diffusive flux of transcription factor molecules to the binding site fluctuates at low input concentration $c$;  in effect the receptor site ``counts'' the number of molecules arriving into its vicinity during a time window $\tau_p$, and this number is of the order $\sim D a c \tau_p$. This argument is conceptually the same as that for the limits to chemoatractant detection in chemotaxis, as discussed by Berg and Purcell  \cite{berg+purcell_77}. It can be shown that this is a theoretical noise floor that cannot be circumvented by using sophisticated ``binding site machinery''   as long as this machinery is contained within a region of linear size $a$ \cite{bialek+setayeshgar_05,bialek+setayeshgar_06}.  For example, cooperative binding  to the promoter or promoters with multiple internal states will modify the binomial switching term, but will leave the diffusion noise unaffected if we express it as an effective noise in transcription factor concentration $\sigma_c$ such that
\begin{equation}
\sigma_{\rm p} = {\bigg |}{{\partial p}\over{\partial c}}{\bigg |} \sigma_c .
\end{equation}

\begin{figure}[t] 
   \centering
   \epsfig{file=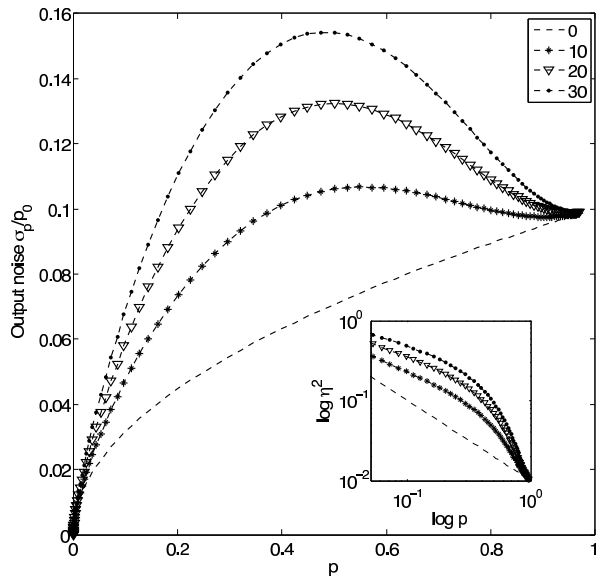,width=0.9\linewidth}
   \caption{Expression noise as a  function of the mean.
The standard deviation of the protein concentration $\sigma_{\rm p}/p_0$ is plotted against the mean protein concentration $\bar{p} = \langle p \rangle/p_0$, from Eq (\ref{eq-model}) with $h=5$. In all cases the output noise term has a strength $\alpha = 0.01$, and the different curves are indexed by the ratio of input noise to output noise $\beta/\alpha = 0, 10, 20, 30$.  In the absence of input noise, the noise level is a monotonic function of the mean, but input noise contributes a peak near the point of half maximal expression ${\bar p} = 0.5$.  
In the inset, we show the same results plotted as a fractional noise variance $\eta_{\rm p}^2$ vs the mean [Eq (\ref{eq-model_eta})], on a logarithmic scale, and we see that the prominent peak has become just an inflection. For most of the dynamic range of means, the contribution of input noise is to increase the fractional variance without substantial changes in the slope of the double--log plot, so that we can confuse input noise with a larger level of output noise, especially if we remember that real data will be scattered due to measurement errors.} 
   \label{f-noise1}
\end{figure}

Although cooperativity does not change the effective concentration noise due to diffusion, it does reduce the relative significance of the switching noise \cite{bialek+setayeshgar_06}.  Since we will discuss a system which is strongly cooperative, in much of what follows we neglect the switching noise term and focus on the output noise and diffusion noise.  Then the generalization to multisite, cooperative regulation is straightforward (see Appendix B).  We expect that  cooperative effects among $h$ transcription factors generate a sigmoidal dependence of expression on the transcription factor concentration, so that
\begin{equation}
\bar{p}=\frac{c^h}{c^h+K_d^h} \label{eq-hill},
\end{equation} 
where $h$ is called the Hill coefficient, and $K_d$ is the concentration required for half maximal activation.  We can invert this relationship to write the concentration $c$, which is relevant for the diffusive noise, as a function of the mean fractional expression level $\bar p$. Substituting back into Eq (\ref{eq-pnoise1}), and neglecting the switching noise, we  obtain
\begin{equation}
\left(\frac{\sigma_{\rm p}}{p_0}\right)^2=\alpha\, \bar{p}+\beta\, \bar{p}^{2-1/h}(1-\bar{p})^{2+1/h}  ,
\label{eq-model}
\end{equation}
where $\alpha$ and $\beta$ are combinations of parameters that measure the strength of the output and diffusion noise, respectively.  If we express the variance in fractional terms, this becomes
\begin{equation}
\eta_{\rm p}^2 = \alpha {1\over {\bar p}} + \beta {\bar p}^{-1/h} (1-{\bar p})^{2+1/h} .
\label{eq-model_eta}
\end{equation}
The global output noise model of Eq (\ref{global_model}) corresponds to $\beta =0$ (no input noise) and $b =\alpha p_0$.  Figure \ref{f-noise1} shows the predicted noise levels for different ratios of output to  input  noise ($\beta/\alpha$).

For very highly cooperative, essentially switch--like systems, we can take the limit $h\rightarrow\infty$ to obtain
\begin{eqnarray}
\left(\frac{\sigma_{\rm p}}{p_0}\right)^2
&=&\alpha\, \bar{p}+\beta\, \bar{p}^{2}(1-\bar{p})^{2}  \label{hinf1}\\
\eta_{\rm p}^2 &=&
 \alpha {1\over {\bar p}} + \beta  (1-{\bar p})^{2} .
\label{hinf2}
\end{eqnarray}
In particular,  if we explore only expression levels well below the maximum (${\bar p} \ll 1$), then the diffusion noise just add a constant $\beta$ to the fractional variance.  Thus, diffusion noise in a highly cooperative system could be confused with a global or even extrinsic noise source.

\section{Signatures of input noise}

Input noise arises from fluctuations in the occupancy of the transcription factor binding sites.  Thus, if we go to very high transcription factor concentrations, where all sites are fully occupied, or to very low concentrations, where the sites are never occupied, the fluctuations must vanish.  These limits correspond, in the case of a transcriptional activator, to maximal and minimal expression levels, respectively.  Thus, the key signature of input noise is that it must be largest at some intermediate expression level, as shown in Fig \ref{f-noise1}.

The claim that many genes have expression noise levels which fit the global output noise model of Eq (\ref{global_model}) would seem to contradict the prediction of  a peak in the noise as a function of the mean.  But if we plot the predictions of the model  with input noise as a fractional variance vs mean, the prominent peak disappears (inset to Fig \ref{f-noise1}).  In fact, over a large dynamic range, the input noise seems just to increase the magnitude of the fractional variance while not making a substantial change in the slope of $\log(\eta_{\rm p}^2)$ vs $\log(\langle p \rangle)$.  Confronted with real data on a system with significant input noise, we could thus fit much of those data with the global output noise model but with a larger value of $b$.  There is, of course, a difference between input and output noise, even when plotted as $\log(\eta_{\rm p}^2)$ vs $\log(\langle p \rangle)$, namely a rapid  drop in noise level as we approach maximal expression.  But this effect is confined to a narrow range, essentially a factor of two in mean expression level.  As we discuss below, there are variety of reasons why this might not have been seen in the data of Ref \cite{bar-even+al_06}.

\begin{figure}
   \centering
 \epsfig{file=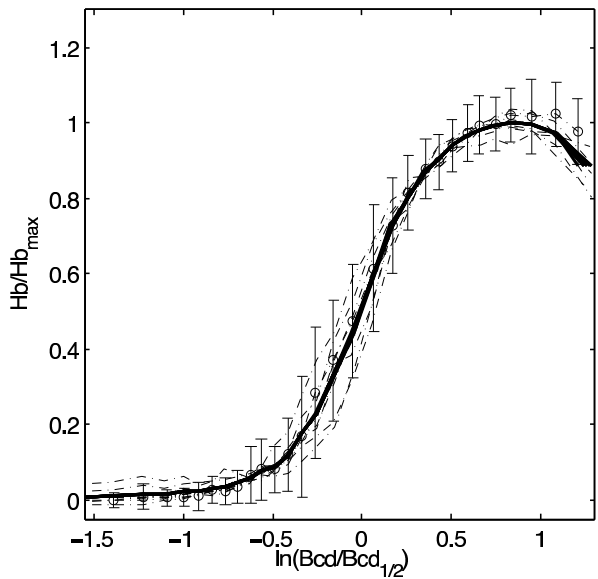,width=0.9\linewidth}
   \caption{The input--output relation for Bicoid regulation of Hunchback expression, redrawn from Ref \cite{gregor+al_06b}. Dashed curves show mean expression levels in  different embryos, thick black line is the mean across all embryos, and points with error bars show the mean and standard deviation of Hb expression at a given Bcd concentration in one embryo.} 
   \label{f-noise2}
\end{figure}

Recent experiments on the precision of gene expression in the early {\em Drosophila} embryo provide us with an opportunity to search for the signatures of input noise \cite{gregor_05,gregor+al_06b}.  The embryo contains a spatial gradient of the protein Bicoid (Bcd), translated from maternal mRNA, and this protein is a transcription factor which activates, among other genes, {\em hunchback}.  Looking along the anterior--posterior axis of the embryo one thus has an array of nuclei that experience a graded range of transcription factor concentrations.  Using antibody staining and image processing methods, it thus is possible to collect thousands of points on a scatter plot of input (Bicoid concentration) vs. output (Hunchback protein concentration); since even in a single embryo there are many nuclei that have the same Bcd concentration, one can examine both the mean Hunchback (Hb) response and its variance; data from Ref \cite{gregor+al_06b} are shown in Fig \ref{f-noise2}.

\begin{figure}[b]  
   \centering
     \epsfig{file=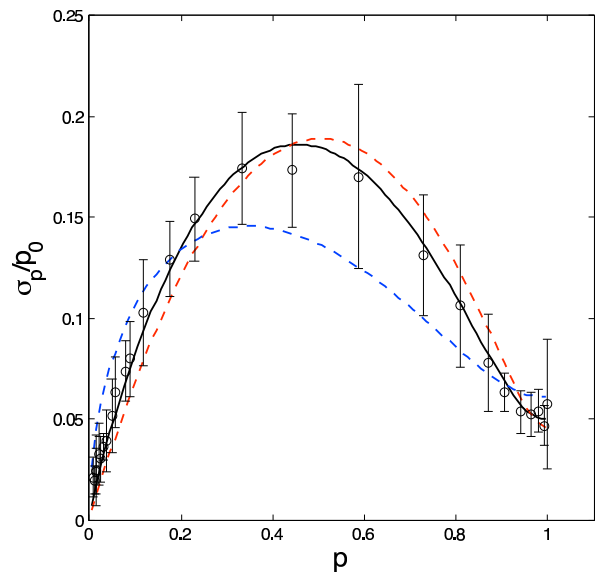,width=0.9\linewidth}
   \caption{Standard deviation of Hunchback expression as a function of the mean (points with error bars), replotted from Ref \cite{gregor+al_06b}.
The black line is a fit of combined output and diffusion noise contributions, from Eq (\ref{eq-model}) with $h=5$, and the dashed red line is with $h\rightarrow\infty$, from Eq (\ref{hinf1}).
In contrast, the dashed blue line is the best fit of combined output and switching noise contributions.  Although both diffusion and switching noise produce a peak at intermediate expression levels, the shapes of the peaks are distinguishable, and the data favor the diffusion noise model.} 
   \label{f-noise3}
\end{figure}

The mean response of Hb to Bcd is fit reasonably well by Eq (\ref{eq-hill}) with a  Hill coefficient $h=5$ \cite{gregor+al_06b}, and in Fig \ref{f-noise3} we replot the noise in this response as a function of the mean.  The peak of expression noise near half maximal expression---the signature of input noise---is clearly visible. More quantitatively, we find that the data are well fit by Eq (\ref{eq-model}) with the contribution from output noise ($\alpha \approx 1/380$) much smaller than that from input noise ($\beta \approx 1/2$).
We also consider the same model with $h\rightarrow\infty$, and this fully switch--like model, although formally still within error bars, systematically deviates from the data. 
Finally we consider a model in which diffusion noise is absent, but we include the switching noise from Eq (\ref{eq-pnoise1}), which generalizes to the case of cooperative binding (see Appendix B).  
Interestingly, this model has the same number of parameters as the diffusion noise model, but does a significantly poorer job of fitting the data.
While the fit can be improved further by adding a small background to the noise, we  emphasize that Eq (\ref{eq-model}) correctly captures the non--trivial shape of the noise curve with only two parameters. Because input noise falls to zero at maximal expression, the sole remaining noise at that point is the output noise, and this uniquely determines the parameter $\alpha$.   The strength of the input noise ($\beta$) then is determined by the height of the noise peak, and there is no further room for adjustment. The {\sl shape} of the peak  is predicted by the theory with no additional parameters, and the different curves in Fig \ref{f-noise3} demonstrate that the data can distinguish among various functional forms for the peak.

Are the parameters $\alpha$ and $\beta$ that fit the Bcd/Hb data biologically reasonable?   The fact that diffusive noise dominates at intermediate levels of expression  ($\beta \gg \alpha$) is the statement that the Hunchback expression level provides a readout of Bcd concentration with a reliability that is close to the physical limit set by diffusional shot noise, as was argued in Ref \cite{gregor+al_06b} based on the magnitude of the noise level and estimates of the relevant microscopic parameters that determine $\beta$.  The dominance of diffusive noise over switching noise presumably is related to the high cooperativity of the Bcd/Hb input/output relation \cite{bialek+setayeshgar_06}.

The parameter $\alpha$ measures the strength of the output noise and thus depends on the absolute number of Hb molecules and on the number proteins produced per mRNA transcript.  If this burst size in the range $R_p \tau_e \sim 1-10$, then our fit predicts the maximum expression level of Hb corresponds to $p_0 = 700-4000$ molecules in the nucleus.  Given the volume of the nuclei at this stage of development ($\sim 140\,\mu{\rm m}^3$; see Refs \cite{gregor+al_06b,gregor+al_06a}), this is a concentration of $8-48\,{\rm nM}$.  Although we don't have independent measurements of the absolute Hunchback concentration, this is reasonable for transcription factors, which typically act in the nanoMolar range \cite{ptashne_92,pedone+al_96,ma+al_96,burz+al_98,winston+al_99,zhao+al_02}, and can be compared with the maximal nuclear concentration of Bcd, which is $55\pm3\,{\rm nM}$ \cite{gregor+al_06b}.  Larger burst sizes would predict larger maximal expression levels, or conversely measurements of absolute expression levels might give suggestions about the burst size for translation in the early {\em Drosophila} embryo.

\section{Discussion}
\begin{figure}[b]
   \centering
     \epsfig{file=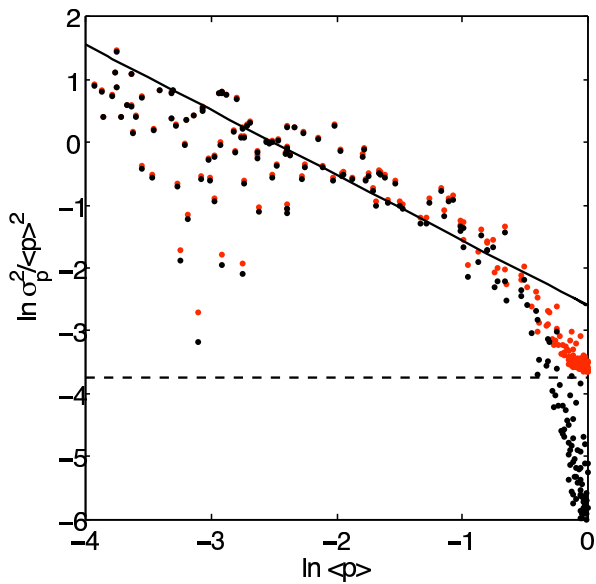,width=0.9\linewidth}
   \caption{Logarithmic plot of fractional variance vs the mean expression level for Hunchback, replotted from Ref \cite{gregor+al_06b}. Each black point represents the noise level measured across nuclei that experience the same Bcd concentration within one embryo, and results are collected from nine embryos.  The solid line shows a fit to $\eta_{\rm p}^2 \propto \langle p\rangle^{-\gamma}$ in the region below half maximal mean expression; we find a good fit, with $\gamma = 1.04$, despite the fact that these data show a clear signature of input noise when plotted in Fig \ref{f-noise3}.
   Dashed line indicates the global noise floor suggested in Ref \cite{bar-even+al_06}, and red points show the raw data with this variance added.  Although the input noise still appears as a  drop in fractional noise level near maximal mean expression, this  now is quite subtle and easily obscured by experimental errors.} 
   \label{f-noise4}
\end{figure}

In the process of transcriptional regulation, the (output) expression level of regulated genes acts as a sensor for the (input) concentration of transcription factors.  The performance of this sensor, and hence the regulatory power of the system, is limited by noise.  While changes in the parameters of the transcriptional and translational apparatus can change the level of output noise, the input noise is determined by the physical properties of the transcription factor and its interactions with the target sites along the genome.  Ultimately, there is a lower bound on this input noise level set by the shot noise in random arrival of the transcription factors at their targets, in much the same way that any imaging process ultimately is limited by the random arrival of photons.

Input and output noise seem to be so different that it is hard to imagine that they could be confused experimentally.  Some of the difficulty, however, can be illustrated by plotting the results from the Bcd/Hb experiments of Ref \cite{gregor+al_06b} in the form which has become conventional in the study of gene expression noise, as a fractional variance vs mean expression level (Fig \ref{f-noise4}).
The signature of input noise, so clear in Fig \ref{f-noise3}, now is confined to a narrow range ($\sim \times 2$) near maximal expression.  In contrast, over more than a decade of expression levels the noise level is a good fit to $\eta_{\rm p}^2 \propto \langle p\rangle^{-\gamma}$, with $\gamma = 1.04$ being very similar to the prediction of the global noise model ($\gamma = 1$) in Eq (\ref{global_model}).    The departures from power--law behavior are easily obscured by global noise sources, experimental error, or by technical limitations that  lead to the exclusion of data at the very highest expression levels, as in Ref \cite{bar-even+al_06}.

The lesson from this analysis of the Bicoid/Hunchback data is that the signatures of input noise are surprisingly subtle.  In this system, however, the behavior near half maximal expression is exactly the most relevant question biologically, since this is where the `decision' is made to draw a boundary, as a first step in spatial patterning.  In other systems, the details of noise in this region of expression levels might be less relevant for the organism, but it is only in this region that different sources of noise are qualitatively distinguishable, as is clear from Fig \ref{f-noise4}.  Thus, unless we have independent experiments to measure some of the parameters of the system, we need experimental access to the full range of expression levels and hence, implicitly, to the full dynamic range of transcription factor concentrations, if we want to disentangle input and output noise.

The early {\em Drosophila} embryo is an attractive model system precisely because the organism itself generates a broad range of transcription factor concentrations, and conveniently arranges these different samples along the major axes of the embryo.  A caveat is that since we don't directly control the transcription factor concentration, we have to measure it.  In particular, in order to measure the variance of the output (Hunchback, in the present discussion) we have to find many nuclei that all have the same input transcription factor (Bicoid) concentration.  Because the mean output is a steep function of the input, errors in the measurement of transcription factor concentration can simulate the effects of input noise, as discussed in Ref \cite{gregor+al_06b}.  Thus, a complete analysis of input and output noise requires not only access to a wide range of transcription factor concentrations, but rather precise measurements of these concentrations.

Why are the different sources of noise so easily confused?  If noise is dominated by randomness in a single step of the translation process, then the number of protein molecules will obey the Possion distribution, and the variance in copy number will be equal to the mean.  But if we can't actually turn measurements of protein level into molecule counts, then all we can say is that the variance will be {\sl proportional} to the mean.  
If the dominant noise source is a single step in transcription, then the number of mRNA transcripts will obey the Poisson distribution, and the variance of protein copy numbers still will be proportional to the mean, but the proportionality constant will be enhanced by the burst size.  The same reasoning, however, can be pushed further back:  if, far from maximal expression, the dominant source of noise is the infrequent binding of a transcriptional activator (or dissociation of a repressor) to its target site, then the variance in protein copy number still will be proportional to the mean.  Thus, the proportionality of variance to mean implies that there is some single rare event that dominates the noise, and by itself doesn't distinguish the nature of this event.

If noise is dominated by regulatory events, then the number of mRNA transcripts should be drawn from a distribution broader than Poisson.  In effect the idea of bursting, which amplifies protein relative to mRNA number variance, applies here too, amplifying the variance of transcript number above the expectations from the Poisson distribution.  Transcriptional bursting has in fact been observed directly \cite{golding+al_05}, although it is not clear whether this arises from fluctuations in transcription factor binding or from other sources.

Previous arguments have made it plausible that input noise is significant in comparison to the observed variance of gene expression \cite{bialek+setayeshgar_05}, and we have shown here that models which assign all of the noise to common factors on the output side are inconsistent with the embedding of gene expression in a regulatory network.  The signatures of input noise seem clear, but can be surprisingly subtle to distinguish in real data.  We have argued that the Bicoid/Hunchback system provides an example in which input noise is dominant, and further that the detailed form of the variance vs mean supports a dominant role for diffusion rather than switching noise.    Although there are caveats, this is consistent with the idea that, as with other critical biological processes \cite{barlow_81,berg+purcell_77,bialek_87,bialek_02},  the regulation of gene expression can operate with a precision limited by fundamental physical principles.

\acknowledgments{We thank CG Callan, JB Kinney, P Mehta, RR de Ruyter van Steveninck, S Setayehsgar, DW Tank, and EF Wieschaus for helpful discussions.
This work was supported in part by the MSREC program of the NSF under Award Number DMR--0213706, by NIH grants  P50 GM071508 and R01 GM077599, 
and by the Burroughs Wellcome Fund Program in Biological Dynamics.}
\appendix
\section{}
We consider a simplified model of regulated gene expression, as schematized in Fig \ref{f-model}:
\begin{eqnarray}
\partial_t c&=&D\nabla^2 c(\vek{x},t) - \dot{n}\, \delta(\vek{x}-\vek{x}_0)+\mathcal{S}-\mathcal{D} \label{eq-dyn0}\\
\dot{n}&=&k_+ c(\vek{x_0},t) (1-n) -k_-n+\xi_n	\label{eq-d1} \label{eq-dyn1}\\
\dot{e}&=&R_en-\tau_e^{-1}e	+ \xi_e \label{eq-dyn2}\\
\dot{p}&=&R_p e-\tau_p^{-1}p + \xi_p. \label{eq-dyn3}
\end{eqnarray}
Equation (\ref{eq-dyn0}) describes the diffusion of the transcription factor that can be absorbed to or released from a binding site on the DNA located at $\vek{x}_0$. These transcription factors are produced at sources $\mathcal{S}$ and degraded at sinks $\mathcal{D}$, which can both be spatially distributed and can also contribute to the noise in $c$. Equation (\ref{eq-dyn1}) describes the dynamics of the binding site occupancy; binding occurs with a second order rate constant $k_+$ and unbinding with a first order rate constant $k_-$, and the  dissociation constant of the site is $K_d=k_-/k_+$. The Langevin term $\xi_n$ induces  stochastic (binomial) switching between 
occupied and empty states of the site.
Equations (\ref{eq-dyn2}) and (\ref{eq-dyn3}) describe the production and degradation of mRNA and protein, respectively, and include Langevin noise terms associated with these birth and death processes.

This seems a good place to note that, while conventional, the assumption that transcription and translation are simple one step processes seems a bit strong. We hope to return to this point at another time.

Our goal is to compute the variance in protein copy number, $\sigma^2_p(\bar{c})$. For simplicity we will assume that the transcription factors are present at a fixed total number in the cell and that they do not decay, $\mathcal{S}=\mathcal{D}=0$. We will see that even with this simplification, where the overall concentration of transcription factors does not fluctuate, we still get an interesting noise contribution from the randomness associated with diffusion in Eq (\ref{eq-dyn0}).

Our basic strategy is to find the steady state solution of the model, and then linearize around this to compute the response of the variables $\{n, e, p\}$ to the various Langevin forces $\{\xi_n , \xi_e , \xi_p\}$.  In the linear approximation, the steady states are also the mean values:
\begin{eqnarray}
c &=& \bar c\\
\langle n \rangle  &=& {{k_+ {\bar c}}\over{k_+ {\bar c} + k_-}} = {{\bar c}\over{{\bar c} + K_d}}\\
{\langle e \rangle} &=& R_e\tau_e \langle n \rangle\\
{\langle p \rangle} &=& R_p\tau_p {\langle e \rangle} = p_0 \langle n \rangle ,
\label{meanp}
\end{eqnarray}
where $p_0 = R_e \tau_e R_p \tau_p$ is the maximum mean expression level.  Notice that what we have called $\bar p = \langle p \rangle /p_0$ in the text is just the mean occupancy, $\langle n \rangle$, of the transcription factor binding site.

Small departures from steady state are written in a Fourier representation:
\begin{eqnarray}
c({\vek x} ,t)  &=& \bar c + \int {{d\omega}\over{2\pi}} \int{{d^3 k}\over{(2\pi)^3}} e^{i{\vek k\cdot \vek x}} e^{-i\omega t} \delta\hat c({\vek k},\omega)\\
n &=& \langle n \rangle + \int {{d\omega}\over{2\pi}} e^{-i\omega t}\delta\hat n(\omega)\\
e &=& {\langle e \rangle} + \int {{d\omega}\over{2\pi}} e^{-i\omega t}\delta\hat e(\omega)\\
p &=& {\langle p \rangle} + \int {{d\omega}\over{2\pi}} e^{-i\omega t}\delta\hat p(\omega).
\end{eqnarray}
Similarly, each of the Langevin terms is written in its Fourier representation,
\begin{equation}
\xi_\mu = \int {{d\omega}\over{2\pi}} e^{-i\omega t}\hat\xi_\mu (\omega) ,
\end{equation}
where $\mu = n, e, p$.

As a first step we use the Fourier representation to solve Eq (\ref{eq-dyn0}) for $\delta c({\bf x_0}, t)$ that we need to substitute into Eq (\ref{eq-dyn1}) for the binding site occupancy:
\begin{eqnarray}
\delta c({\bf x_0}, t) &=& \int{{d\omega}\over{2\pi}} e^{-i\omega t}\delta \tilde c ({\bf x_0},\omega)\\
\delta \tilde c({\bf x_0},\omega)&=&i\omega\delta \hat{n}(\omega) \int\frac{d^3 k}{(2\pi)^3}\frac{1}{-i\omega+D|\vek{k}|^2} \label{kint}	\\
&=&\frac{i\omega \delta\hat{n}(\omega) }{\pi D a} .\label{eq-cpl1}
\end{eqnarray}
The integral over $\vek k$ in Eq (\ref{kint}) is divergent at large $|{\vek k}|$ (ultraviolet).  This arises, as explained in Ref \cite{bialek+setayeshgar_05}, because we started with the assumption that the binding reaction occurs at a point---the delta function in Eq (\ref{eq-dyn0}).  In fact our description needs to be coarse grained on a scale corresponding to the size of the binding site, so we introduce a cutoff so that
$|{\vek k}| \leq k_\mathrm{max}={2\pi}/{a}$, where $a$ is the linear size of the binding site.

Linearizing Eq (\ref{eq-dyn1}) for the dynamics of the site occupancy, we have
\begin{widetext}
\begin{equation}
-i\omega\delta \hat n (\omega) = -(k_+ \bar c + k_- ) \delta\hat n (\omega ) 
+k_+ (1-\langle n \rangle ) \delta\tilde c ({\bf x_0},\omega ) + \hat\xi_n(\omega ).
\end{equation}
Substituting our result for $\delta \tilde c(\vek{x}_0,\omega)$   from Eq (\ref{eq-cpl1}),
we find
\begin{eqnarray}
-i\omega\delta \hat n (\omega)&=&
 -(k_+ \bar c + k_- ) \delta\hat n (\omega ) 
+k_+ (1-\langle n \rangle )\frac{i\omega \delta\hat{n}(\omega) }{\pi D a} + \hat\xi_n(\omega )\\
-i\omega \left[ 1 + {{k_+ (1-\langle n \rangle )}\over{\pi D a}}\right] \delta \hat n (\omega)
&=& -(k_+ \bar c + k_- ) \delta\hat n (\omega ) + \hat\xi_n(\omega )\\
\delta \hat n (\omega ) &=&
{{\hat\xi_n (\omega )}\over{ -i\omega ( 1 + \Sigma ) 
+ (k_+ \bar c + k_- )}}\label{lin1}
\end{eqnarray}
where $\Sigma = k_+ (1-\langle n \rangle )/(\pi D a)$.  The linearization of Eqs (\ref{eq-dyn2}) and (\ref{eq-dyn3}) takes the form
\begin{eqnarray}
-i\omega \delta\hat e(\omega) &=& -{1\over{\tau_e}} \delta\hat e(\omega ) + R_e \delta\hat n(\omega ) + \hat\xi_e (\omega)\label{lin2}\\
-i\omega \delta\hat p(\omega) &=& -{1\over{\tau_p}} \delta\hat p(\omega ) + R_p \delta\hat e(\omega ) + \hat\xi_p (\omega)\label{lin3}
\end{eqnarray}
Each Langevin term is independent, and each frequency component $\omega$ is correlated only with the component at $-\omega$, defining the noise power spectrum
$\langle \hat \xi_\mu (\omega ) \hat \xi_\mu (-\omega ') \rangle
= 2\pi\delta(\omega -\omega ') {\cal N}_\mu (\omega )$ for $\mu = n,e, p$.  Solving the three linear equations, Eqs (\ref{lin1}--\ref{lin3}), we can find the power spectrum of the protein copy number fluctuations,
\begin{equation}
\mathcal{S}_{p}(\omega)
=\frac{{\cal N}_p}{\omega^2+1/\tau_p^2}
+
R_p^2\frac{{\cal N}_e}{(\omega^2+1/\tau_p^2)(\omega^2+1/\tau_e^2)}
+
R_p^2R_e^2\frac{{\cal N}_n}{(\omega^2+1/\tau_p^2)(\omega^2+1/\tau_e^2)[(1+\Sigma)^2\omega^2+1/\tau_c^2]},
\label{bigspec}
\end{equation}
\end{widetext}
where $1/\tau_c = k_+ \bar c + k_-$.  This form has a very intuitive interpretation: each Langevin  term represents a noise source; as this noise propagates from the point where it enters the dynamical system to the output, it is subjected both to gain of each successive stage (prefactors $R$), and to  filtering by factors of $\mathcal{F}_\tau=(\omega^2+1/\tau^2)^{-1}$.

The total variance in protein copy number is given by an integral over the spectrum,
\begin{equation}
\langle (\delta p)^2\rangle \equiv \sigma_p^2  = \int {{d\omega}\over{2\pi}}\mathcal{S}_p (\omega ) ,\label{sigma=int}
\end{equation}
and the noise power spectra of the Langevin terms associated with the mRNA and protein dynamics have the simple forms ${\cal N}_e (\omega ) = 2R_e \langle n \rangle$ and ${\cal N}_p(\omega ) = 2R_p\langle e \rangle$, respectively.  The spectrum ${\cal N}_n (\omega )$ is more subtle.  One way to derive it is to realize that since there is only one binding site and this site is either occupied or empty, the total variance of $\delta n$ must be given by the binomial formula,
\begin{equation}
\langle (\delta n)^2 \rangle = {\langle n \rangle} (1 - \langle n \rangle ).
\label{binomial}
\end{equation}
Starting with Eq (\ref{lin1}) and the analog of Eq (\ref{sigma=int}), we can use this condition to set the magnitude of ${\cal N}_n$.  Alternatively, we can use the fact that binding and unbinding come to equilibrium, and hence the fluctuations in $n$ are a form of thermal noise, like Brownian motion or Johnson noise, and hence the spectrum ${\cal N}_n$ is determined by the fluctuation--dissipation theorem \cite{bialek+setayeshgar_05}.  The result is that
\begin{equation}
{\cal N}_n = {2\over {\tau_c}} (1+\Sigma ) \langle n \rangle ( 1-\langle n \rangle ) .
\end{equation}

For simplicity we consider the case where the protein lifetime $\tau_p$ is long compared with all other time scales in the problem.  Then we can approximate Eq (\ref{bigspec}) as
\begin{equation}
\mathcal{S}_p(\omega )\approx
{1\over{\omega^2 + 1/\tau_p^2}}\left[
{\cal N}_p + (R_p\tau_e)^2 {\cal N}_e + (R_p\tau_e R_e\tau_c)^2{\cal N}_n\right] .
\end{equation}
Substituting the forms of the individual noise spectra ${\cal N}_\mu$ and doing the integral over $\omega$ [Eq (\ref{sigma=int})], we find the variance in protein copy number
\begin{eqnarray}
\sigma_p^2 &=& \tau_p [R_p \langle e \rangle + (R_p\tau_e)^2 R_e\langle n \rangle]\nonumber\\
&& + {{\tau_p}\over {\tau_c}}(R_p\tau_e R_e\tau_c)^2 (1+\Sigma) \langle n \rangle (1 - \langle n \rangle ) .
\label{var1}
\end{eqnarray}
We notice that the first term in this equation is $R_p\tau_p\langle e \rangle$, which is just the mean number of proteins $\langle p\rangle$ from Eq (\ref{meanp}).  The second term
\begin{eqnarray}
\tau_p (R_p\tau_e )^2 R_e\langle n \rangle &=& R_p\tau_p (R_e\tau_e\langle n \rangle ) (R_p\tau_e)\\
&=& R_p\tau_p \langle e \rangle (R_p\tau_e)\\
& =& R_p\tau_e \langle p\rangle .
\end{eqnarray}
Thus, the first two terms together contribute $(1+R_p\tau_e)\langle p\rangle$ to the variance, and this corresponds to the output noise term in Eq (\ref{eq-model}).

The third term in Eq (\ref{var1}) contains the contribution of input noise to the variance in protein copy number.  To simplify this term we note that the steady state of Eq (\ref{eq-dyn1}) is equivalent to
\begin{equation}
k_+ \bar c (1-\langle n\rangle ) = k_- \langle n \rangle.
\label{k+vsk-}
\end{equation}
Thus we can write
\begin{eqnarray}
{1\over {\tau_c}} &\equiv& k_+ \bar c + k_- \\
&=& k_-\left[ {{\langle n \rangle}\over{1-\langle n\rangle}} +1\right]
= {{k_-}\over{1-\langle n \rangle}} .
\end{eqnarray}
The term we are interested in is
\begin{widetext}
\begin{eqnarray}
{{\tau_p}\over {\tau_c}}(R_p\tau_e R_e\tau_c)^2 (1+\Sigma) \langle n \rangle (1 - \langle n \rangle )
&=& (R_p\tau_p R_e\tau_e)^2 {{\tau_c}\over{\tau_p}} (1+\Sigma) \langle n \rangle (1 - \langle n \rangle )\\
&=& p_0^2 {1\over{k_-\tau_p}} (1+\Sigma ) \langle n \rangle(1-\langle n \rangle)^2 \\
&=& p_0^2 {{\langle n \rangle(1-\langle n \rangle)^2}\over{k_-\tau_p}}
+ p_0^2 {1\over{k_-\tau_p}} {{k_+ (1-\langle n \rangle )}\over{\pi D a}}\langle n \rangle(1-\langle n \rangle)^2 \\
&=& p_0^2 {{\langle n \rangle(1-\langle n \rangle)^2}\over{k_-\tau_p}}
+ p_0^2 {{\langle n \rangle^2 (1-\langle n \rangle )^2}\over{\pi D a \bar c \tau_p}} ,
\label{var_end}
\end{eqnarray}
\end{widetext}
where in the last step we once again use Eq (\ref{k+vsk-}) to rewrite the ratio $k_+/k_-$ in terms of $\langle n \rangle$.  We recognize the two terms in this result as the switching and diffusion terms in Eq (\ref{eq-model}).

\section{}

To generalize this analysis of noise to cooperative interactions among transcription factors it is useful to think more intuitively about the two terms in Eq (\ref{var_end}), corresponding to switching and diffusion noise.  Consider first the switching noise.

We are looking at a binary variable $n$ such that the number of proteins is $p_0 n$.  The total variance in $n$ must be $\langle (\delta n)^2\rangle = \langle n \rangle (1- \langle n \rangle )$ [Eq (\ref{binomial})].  This noise fluctuates on a time scale $\tau_c$, so during the lifetime of the protein we see $N_s = \tau_p/\tau_c$ independent samples. The current protein concentration is effectively an average over these samples, so the effective variance is reduced to 
\begin{equation}
\langle (\delta n)^2\rangle_{\rm eff} = 
{1\over{N_s}} \langle n \rangle (1- \langle n \rangle ) = 
{{\tau_c}\over{\tau_p}} \langle n \rangle (1- \langle n \rangle ) .
\end{equation}
Except for the factor of $p_0$ that converts $n$ into $p$, this is the first term in Eq (\ref{var_end}).

Now if $h$ transcription factors bind cooperatively, we can still have two states, one in which transcription is possible and one in which it is blocked.  For the case of activation, which we are considering here, the active state corresponds to all binding sites being filled, and so the rate at which the system leaves this state, $k_-$, shouldn't depend on the concentration of the transcription factors.  The rate at which the system enters the active state does depend on concentration, but this doesn't matter, because with only two states we must always have an analog of Eq (\ref{k+vsk-}), which allows us to eliminate the ''on rate'' in favor of $k_-$ and $\langle n \rangle$.  The conclusion is that the first term in Eq (\ref{var_end}), corresponding to switching noise, is unchanged by cooperativity as long as the system is still well approximated as having just two states of transcriptional activity that depend on the potentially many more states of binding site occupancy.

For the diffusion noise term we use the ideas of Refs \cite{berg+purcell_77,bialek+setayeshgar_05,bialek+setayeshgar_06}.  Diffusion noise should be thought of as an effective noise in the measurement of the concentration $c$, with a variance
\begin{equation}
{{\sigma_c^2}\over{\bar c^2}} \sim {1\over{\pi D a \bar c \tau_p}} ,
\label{DacT}
\end{equation}
where again we identify the protein lifetime as the time over which the system averages.
For the system with a single binding site,
\begin{equation}
\langle n \rangle = {{\bar c}\over{\bar c + K_d}} ,
\end{equation}
so that 
\begin{equation}
{{\partial \langle n \rangle}\over{\partial c}}  = 
{1\over{\bar c}} \langle n \rangle ( 1 - \langle n \rangle ) .
\end{equation}
The noise in concentration, together with this sensitivity of $n$ to changes in the concentration, should contribute a noise variance
\begin{equation}
\langle (\delta n)^2\rangle_{\rm eff} =  {\bigg |} {{\partial \langle n \rangle}\over{\partial c}}
{\bigg |}^2 \sigma_c^2 = {{\langle n \rangle^2 ( 1 - \langle n \rangle )^2}\over{\pi D a \bar c \tau_p}} .
\end{equation}
This is (up to the factor of $p_0$) the second term in Eq (\ref{var_end}).  Now the generalization to cooperative interactions is straightforward.  If we have
\begin{equation}
\langle n \rangle = {{\bar c}^h\over{\bar c^h + K_d^h}} ,
\end{equation}
then
\begin{equation}
{{\partial \langle n \rangle}\over{\partial c}}  = 
{h\over{\bar c}} \langle n \rangle ( 1 - \langle n \rangle ) .
\end{equation}
Since the effective noise in concentration is unchanged \cite{bialek+setayeshgar_06}, the only effect of cooperativity is to multiply the second term in Eq (\ref{var_end}) by a factor of $h^2$.  

Thus, in the expression [Eq (\ref{eq-model})] for the variance of protein copy number, cooperativity has no effect on the switching noise by actually increases the diffusion noise by a factor of $h^2$.  When written as a function of the mean copy number and the  transcription factor concentration, this leaves the functional form of the variance fixed, only changing the coefficients.  The overall effect it to make the contribution of diffusion noise more important.  One way to say this is that, when we refer the noise in copy number back to the input, cooperativity causes the equivalent concentration noise to become closer to the limit Eq (\ref{DacT}) set by diffusive shot noise \cite{bialek+setayeshgar_06}.

Reference \cite{gregor+al_06b} also considers the possibility that noise is reduced by averaging among neighboring nuclei. This does not change the form of any of the noise terms, but does change the microscopic interpretation of the coefficients $\alpha$ and $\beta$.  For example, averaging for a time $\tau_p$ over $N$ nuclei is equivalent to having one nucleus with an averaging time $N\tau_p$.

\end{document}